\documentclass[prd,twocolumn,showpacs,letterpaper,nofootinbib,superscriptaddress,floatfix]{revtex4}
\usepackage{graphicx}
\usepackage{times}
\usepackage{amsmath}

\def\thercsid{\relax}
\def\rcsid#1{\def\next##1#1{\def\thercsid{##1}}\next}
\rcsid$Id: paper.tex,v 1.38 2009/01/21 12:12:36 jsread Exp $

\renewcommand{\today}{\number\day\space\ifcase\month\or
  January\or February\or March\or April\or May\or June\or
  July\or August\or September\or October\or November\or December\fi
  \space\number\year}

\begin{document}
\title{Measuring the neutron star equation of state with gravitational wave observations}
\author{Jocelyn S. Read}
\affiliation{Max-Planck-Institut f\"{u}r Gravitationsphysik, Albert-Einstein-Institut, Golm, Germany}
\author{Charalampos Markakis}
\affiliation{Department of Physics, University of Wisconsin--Milwaukee, P.O. Box 413, Milwaukee, WI 53201, USA}
\author{Masaru Shibata}
\affiliation{Graduate School of Arts and Sciences, University of Tokyo, Komaba, Meguro, Tokyo 153-8902, Japan}
\author{K\=oji Ury\=u}
\affiliation{Department of Physics, University of the Ryukyus, 1 Senbaru, Nishihara, Okinawa 903-0213, Japan}
\author{Jolien D. E. Creighton}
\affiliation{Department of Physics, University of Wisconsin--Milwaukee, P.O. Box 413, Milwaukee, WI 53201, USA}
\author{John L. Friedman}
\affiliation{Department of Physics, University of Wisconsin--Milwaukee, P.O. Box 413, Milwaukee, WI 53201, USA}
%\date[\relax]{RCS \thercsid; compiled \today}
\begin{abstract}%
We report the results of a first study that uses numerical
simulations to estimate the accuracy
with which one can use gravitational wave observations
of double neutron star inspiral to measure parameters of the neutron-star
equation of state.  The simulations use the evolution and initial-data
codes of Shibata and Ury\=u to compute the last several orbits and the
merger of neutron stars, with matter described by a parametrized
equation of state.  Previous work suggested the use of an effective cutoff
frequency to place constraints on the equation of state. We find,
however, that greater accuracy is obtained by measuring departures
from the point-particle limit of the gravitational waveform produced
during the late inspiral.  As the stars approach their final plunge
and merger, the gravitational wave phase accumulates more rapidly
for smaller values of the neutron star compactness (the ratio of
the mass of the neutron star to its radius). 
We estimate that realistic equations of state will lead to gravitational
waveforms that are distinguishable from point particle inspirals at an
effective distance (the distance to an optimally oriented and located system
that would produce an equivalent waveform amplitude) of 100\,Mpc or less.  As Lattimer and
Prakash observed, neutron-star radius is closely tied to the pressure
at density not far above nuclear. Our results suggest that broadband gravitational
wave observations at frequencies between 500 and 1000\,Hz will constrain
this pressure, and we estimate the accuracy with
which it can be measured. Related first estimates of radius measurability
show that the radius can be determined to an accuracy of $\delta R \sim
1$\,km at 100\,Mpc. 
% Because these measurements make use of the finite-size effects on the late inspiral
% rather than the oscillations of a quasi-stable final neutron star remnant,
% careful numerical treatments of temperature, neutrino transport, differential rotation,
% magnetic fields, are not important to the qualitative results
% presented here.  
\end{abstract} \maketitle

\section{Introduction}

Gravitational wave observations can potentially measure properties of
neutron star equations of state (EOS) by measuring departures from the
point-particle approximation to the gravitational waveform produced
during the late inspiral phase of binary neutron star coalescence.  We
examine here the accuracy with which detectors with the sensitivity of
Advanced LIGO can extract from inspiral waveforms an EOS
parameter associated with the stiffness of the neutron star EOS
above nuclear density.  

To do this, we use a set of numerical
simulation waveforms produced by varying the EOS used to model the neutron
star matter.  The signal analysis focuses on the late inspiral, as the
radius of the orbit $r$ approaches the neutron star radius $R$. The orbital
dynamics in this region will depend on the radius and internal structure of
the neutron star, which in turn depend on the EOS\@. We also estimate the
accuracy with which neutron star radii, closely linked to the EOS parameter
varied, can be extracted. This is a preliminary study, using a first set of
multi-orbit binary neutron star waveforms. Subsequent work will use an
improved AMR code with higher resolution to obtain higher accuracy and to
more fully explore the EOS parameter space.

The study of radius and EOS effects on gravitational wave
inspiral was first aimed at questions of detectability
\cite{irrotBNS, irrotBNS2}, showing that the tidal effects would
only affect phase evolution at the end of the inspiral and that point
particle waveforms could be used for template-based detection in LIGO%
\footnote{However with the increased sensitivity of Advanced LIGO, the
contribution to phase evolution from tidal effects may affect the waveform 
during the early, low-frequency part of the inspiral
\cite{FlanaganHinderer2007}.}. 
Subsequently, the competing effects of relativity and finite size at the
end of binary neutron star inspiral have been studied with an array of
different approximations, yielding estimates of the
gravitational wave spectrum which depart from that of a point particle
inspiral starting somewhere between 500\,Hz and over 1000\,Hz
\cite{ LaiWiseman,ZhugeEt1996, RasioShapiro1999, BNSinitial1,FaberEt2002,
FaberEt2004, bejger2005qe, gondek2007eos}.

The study of EOS signature on gravitational waveforms has often focused on
the merger and coalescence phases of the waveform
\cite{Shibata:2005ss,Shibata:2002jb,oechslinjanka2007}, above the frequency of
the last orbit.  These frequencies are higher than the sensitive band
of ground-based interferometers like Advanced LIGO, except in carefully
tuned high frequency narrow-band configurations.  Early work on
measurability of finite size effects \cite{hughes2002tun} used a model of
point-particle inspiral truncated at such a frequency, which could be
sought with narrow-band tuning. 

The characteristic frequency where EOS effects become important for the
gravitational waveform is often estimated via an innermost stable circular
orbit or Roche lobe overflow in a quasiequilibrium approximation. With many
such estimates above 1000\,Hz \cite{gondek2007eos,
BNSinitial1, BNSinitial5}, some summaries of neutron star radius
measurability with gravitational waves have assumed
that EOS dependence in double neutron-star binaries is unlikely to be
detected with Advanced LIGO~\cite{baumgarte2008}.

However, for quantitatively studying the late inspiral and merger phases of
binary neutron stars, numerical relativity is required.
Until quite recently, there has been no general relativistic
simulation that quantitatively clarifies the inspiral to merger
phases primarily because of limitation of the computational
resources, although a number of simulations have been done for a
qualitative study~
\cite{Shibata:1999hn,Shibata:1999wm,
ShibataTaniguchi2006,Shibata:2002jb,Shibata:2003ga,Shibata:2005ss,Miller:2003vc,Marronetti:2003hx,Liu:2008xy}.
The crucial drawback in the previous works was that the simulations were
short-term for the inspiral phase; the inspiral motion of the neutron stars
was followed only for about one orbit, and the gravitational wave spectrum
determined only above ~1 kHz
(but see \cite{Yamamoto:2008js,Baiotti:2008ra}).
Furthermore, the simulations were usually started with a quasi-circular
state in which the approaching velocity between two neutron stars is
assumed to be zero and thus the eccentricity is not exactly zero. Thus, in
the previous studies, the radial velocity at the onset of merger is not
correctly taken into account and the non-zero eccentricity could play an
unfavorable role.

In this paper, by examining longer evolutions of binary
neutron stars, the early relaxation from the quasi-circular state can be
removed.  By comparing the numerical inspiral waveform to that
of point particles, we confirm that the frequency evolution in the late
inspiral differs from the point particle case, accumulating phase more
quickly in the final orbits and merging at earlier times. This effect
depends systematically on the EOS and resultant radius of the
neutron stars simulated, and for large variations in EOS and radius the effect
is larger than estimates of error in the numerical waveform.

We calculate the signal strength of this
difference in waveform using the sensitivity curves of commissioned and
proposed gravitational wave detectors, and find that there is a measurably
different signal at reasonable distances from the inspirals of binary
neutron stars with different EOS\@. This leads to a first quantitative
estimate of the measurability of EOS with Advanced LIGO.

We also note that the broadband configuration of Advanced LIGO does as well
or better as narrow band configurations in detecting the difference between
neutron star EOS\@.  An improved understanding of the sensitivity of different
gravitational wave detector configurations to neutron star structure will
be essential for the design of next-generation detectors for gravitational
wave astrophysics. 

\section{Equations of state} \label{sec:modeleos}
\begin{figure}[!htb]
\caption[EOS for numerical evolution]{
\label{fig:candidates}Initial choices of EOS for numerical evolution 
compared to the set of tabled EOS considered in \cite{pppp}.
Candidates are labelled in order of increasing softness: 2H, H, HB, B, 2B.}
\begin{center}
\includegraphics[height=80mm]{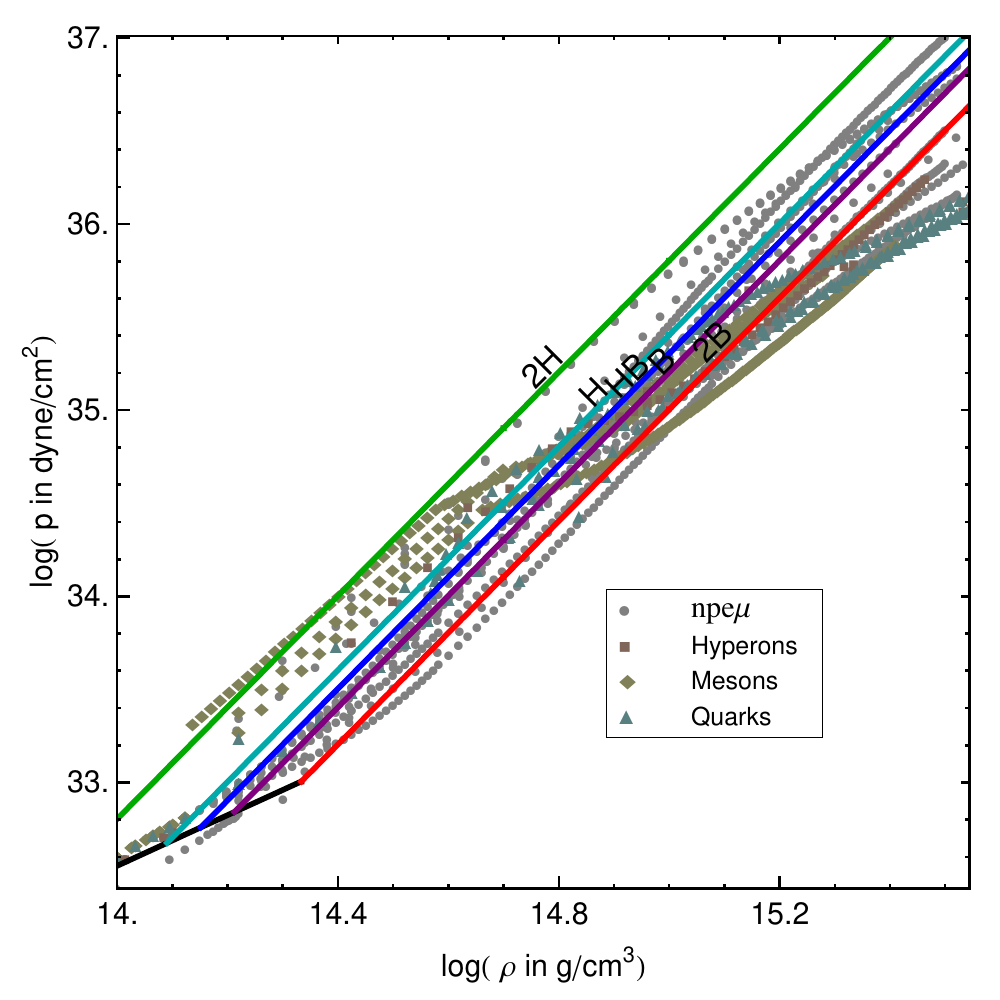} 
\end{center}
\end{figure}

We choose EOS based on work done in \cite{pppp} to 
develop a parameterized EOS that accurately reproduces features of
realistic EOS\@. Systematic variation of the EOS parameters
allows us to determine which properties significantly affect the gravitational radiation
produced, and thus can be constrained with sufficiently strong
gravitational wave detections.

The models chosen for this study use a variation of one EOS parameter in the
neutron star core. The EOS pressure $p$ is specified
as a function of rest mass density%
\footnote{Rest mass density $\rho = m_{\text{b}} n$ is proportional to
number density $n$ with the mass per baryon, if the matter were to be
dispersed to infinity, fixed to be $m_{\text{b}} = 1.66\times
10^{-24}$\,g.}. Rest mass density $\rho$, and the resulting energy density $\epsilon$ are then
determined by the first law of thermodynamics. We use
piecewise polytropic EOS, of the form
\begin{equation} \label{eq:eos}
p(\rho) = K_i \rho^{\Gamma_i}, \quad
d\frac\epsilon\rho = -p\,
d\frac1\rho, 
\end{equation} 
in a set of intervals $\rho_i \leq \rho \leq \rho_{i+1}$ in rest mass
density, with $\epsilon / \rho \to 1$ as $\rho \to 0$.

A fixed crust EOS models the behavior of matter down to 
$10^{12}$\,g\,cm$^{-3}$; the numerical simulations considered do not resolve
densities below this.  The crust is modelled with a single polytrope region,
fitted to tabulated crust EOS, for the region above neutron drip.  The
polytrope has $\Gamma_{\text{crust}} = 1.3569$, with $K_{\text{crust}}$
chosen so that $p/c^2= 1.5689\times10^{31}$\,dyn\,cm$^{-2}$ when
$\rho=10^{13}$\,g\,cm$^{-3}$. The core EOS is constructed independently
of crust behavior, and the dividing density between the crust and core
varies by EOS.

In \cite{pppp} it was found that three zones within the core are
needed to accurately model the full range of proposed EOS models; however in this paper
we will consider only one core zone, described by just one polytropic EOS\@.
We vary the core EOS with an overall pressure shift $p_1$,
specified at the fiducial density $\rho_1 =
5.0119\times10^{14}$\,g\,cm$^{-3}$, while holding the adiabatic index in
all regions of the core fixed at $\Gamma=3$.  
While only a subset of realistic EOS are well-approximated by a single core
polytrope, reducing the EOS considered to this single-parameter family
allows us to estimate parameter measurability with a reasonable
number of simulations.

After fixing the core adiabatic index, increasing the overall pressure
scale $p_1$ produces a family of neutron stars with progressively
increasing radius for a given mass; the $p_1$ parameter was chosen in part
because Lattimer and Prakash \cite{lattprak} found that pressure at one to
two times nuclear density is closely tied to neutron star radius, with $R
\propto p_1^{1/4}$.  The radius is less sensitive to variation of the
adiabatic index in the neutron star core, for reasonable adiabatic indices
\cite{pppp}. 

An important element of future work will be incorporating additional
variations of the EOS within the core. This could involve additional models of
varying adiabatic index around a fixed $p_1$, as well as multiple
piecewise-polytrope zones within the core or EOS parameters yielding
neutron stars of the same $R$ but different internal structure.  Such work
would yield insight into the relative size and correlation of effects on
the orbital evolution due to the stellar radius and internal
structure.
%\begin{table}[!htb]
\begin{table}[t]
\caption[Properties of EOSs for numerical simulation]{Properties of
initial EOS\@. These range from the ``softest''
EOS at the top, which results in a prompt collapse to
a black hole upon merger, to the ``hardest'' (or ``stiffest'') at the
bottom.  Model HB is considered a typical EOS\@.  The pressure
$p_1$, which is the pressure at density
$\rho_1=5\times10^{14}\,\text{g}\,\text{cm}^{-3}$, determines the polytropic
EOS for the neutron star core; all candidates have $\Gamma=3$.
Radius $R$ and compactness $GM/c^2R$ are those of a single isolated $M=1.35\,M_\odot$ TOV star where radius is measured in Schwarzschild-like coordinates.
An astrophysically important EOS-dependent parameter is the maximum neutron
star mass, $M_{\text{max}}$, which is given in the fifth column.
\label{tab:modprop}}
\begin{center}
\begin{tabular}{lcccc}\hline
Model & $\log_{10} p_1\,[\mbox{dyn}\,\mbox{cm}^{-2}]$ & $R\,[\mbox{km}]$
& $GM/c^2R$ & $M_{\text{max}}\,[M_\odot]$\\
\hline\hline
2H & 34.90 & 15.2 & 0.13 & 2.83\\
H  & 34.50 & 12.3 & 0.16 & 2.25\\
HB & 34.40 & 11.6 & 0.17 & 2.12\\
B  & 34.30 & 10.9 & 0.18 & 2.00\\
2B & 34.10 & \phantom{0}9.7 &0.21 & 1.78 \\
\hline
\end{tabular}
\end{center}
\end{table}

The first models were chosen with EOS that ``bracket'' the range of
existing candidates, seen in Fig.~\ref{fig:candidates}. These models
are HB with $p_1=10^{34.40}\,\text{dyn}\,\text{cm}^{-2}$, a standard
EOS; 2H with $p_1=10^{34.90}\,\text{dyn}\,\text{cm}^{-2}$, a stiff
EOS; 2B, with $p_1/c^2=10^{34.10}\,\text{dyn}\,\text{cm}^{-2}$, a soft
EOS\@.  Additional models B with
$p_1/c^2=10^{34.30}\,\text{dyn}\,\text{cm}^{-2}$ and H with
$p_1/c^2=10^{34.50}\,\text{dyn}\,\text{cm}^{-2}$, were chosen with
small shifts in parameter from HB to better estimate local parameter
dependence of the waveform.

%Although the set of models is explicitly constructed by varying the single
%parameter $p_1$,  the resulting neutron star models also have systematic
%variation in the radius $R$, as seen in Table \ref{tab:modprop}. They can
%also be considered a one-parameter family of varying $R$; in fact, $p_1$
%was chosen as an EOS parameter as Lattimer and . The
%parameter estimation analysis is done for both $p_1$ and $R$. 

%We expect the structure of $\sim 1.4 M_\odot$ neutron
%stars to depend most strongly on the value of $p_1$, based on both the
%empirical formula of Lattimer and Prakash \cite{lattprak} describing
%radius as a function of pressure at a fiducial density at some $\rho \simeq
%2.7$ to $5 \times10^14$\,g\,cm$^{-3}$, and the piecewise polytrope equation
%of state study described in Chapter \ref{ch:ppp} showing moment of inertia
%of $1.388 M_\odot$ neutron stars is largely dependent on $p_1$.

\section{Numerical methods}

For each EOS considered, we simulate the late inspiral and
merger of a binary neutron star system. For this study, we fix the
gravitational mass of each neutron star in the binary to $1.35 M_\odot$, an
average value for pulsars observed in binary systems
\cite{ThorsettChakrabarty,DoublePulsar2}. We expect the significance of
tidal effects in this configuration to be fairly representative of tidal
effects over the relatively narrow range of masses and mass ratios expected in
astrophysical binary neutron star system.

\subsection{Initial data}

\begin{table*}
\caption{Quantities of each initial data set for
irrotational binary neutron stars.  
Each star has a baryon number $M_0$ equal to that of an isolated
$M=1.35\,M_\odot$ star. The ADM mass $M_\text{ADM}$ of the initial slice
includes the binding energy, $J$ is the total angular momentum of the
initial slice. The binary compactness $C_0$ is defined
by $C_0=(\Omega M_{\text{ADM}} Gc^{-3})^{2/3}$.}
\label{tab:initial}
\begin{tabular}{lcccccc}
\hline
Model&$\rho_{\text{max}}\,[\mbox{g}\,\mbox{cm}^{-3}]$ &$M_0 \,[M_\odot]$
&$M_{\text{ADM}}
\,[M_\odot]$& $cJ / ( G M_{\text{ADM}}^2)$&$\Omega / 2\pi$ [Hz]& $C_0$ \\
\hline\hline
2H &$3.73196\times10^{14}$ &1.45488 & 2.67262 & 0.993319 & 324.704 &
$8.96966\times10^{-2}$\\
H  &$7.02661\times10^{14}$ &1.48385 & 2.67080 & 0.989524 & 321.468
&$8.90593\times10^{-2}$\\
HB &$8.27673\times10^{14}$ &1.49273 &2.67290 &0.995361 & 309.928 &
$8.69582\times10^{-2}$\\
B &$9.77811\times10^{14}$ &1.50247 & 2.67290 & 0.992638 & 314.170&
$8.77522\times10^{-2}$\\
2B &$1.38300\times10^{15}$ &1.52509 & 2.67229 & 0.987681 &321.170 &
$8.90375\times10^{-2}$\\
\hline
\end{tabular}
\end{table*}

Conformally flat initial data is generated by constructing a
quasiequilibrium sequence of irrotational neutron stars in binary system,
following the methods of
\cite{BNSinitial,BNSinitial1,BNSinitial2,BNSinitial3,BNSinitial4,BNSinitial5}.
%As in previous work, we assume equal-mass binaries, and
As in previous work, we assume irrotational flow fields, neglecting spin of
the neutron stars.  This assumption is based on the estimation of
negligible tidal spin-ups  \cite{irrotBNS,irrotBNS2}.  The parameterized
EOS of Eq.~(\ref{eq:eos}) is incorporated in the code to
solve for initial data with a conformally flat spatial geometry, using the
Isenberg-Wilson-Mathews formulation \cite{IWM1,IWM2} coupled to the
neutron star matter equation consistently.  Each of the stars has a baryon
number equal to that of an isolated  star with gravitational mass
$M=1.35\,M_\odot$. The initial data for the full numerical  simulation is
taken from the quasi-equilibrium configuration at a separation such that $
\sim 3$ orbits remain before merger. Relevant quantities of the initial
configurations for each parameterized EOS are presented in Table~\ref{tab:initial}.

\subsection{Numerical evolution}
The Einstein equations are evolved with the original version of the 
Baumgarte-Shapiro-Shibata-Nakamura formulation~
\cite{Shibata:1995we,bs99} in which we evolve the conformal factor, 
$\varphi=(\ln\gamma)/12$, the trace $K$ of the extrinsic curvature,
the conformal three metrics,
$\tilde{\gamma}_{ij}\equiv\gamma^{-1/3}\gamma_{ij}$, the tracefree part of the
extrinsic curvature, $\tilde{A}_{ij}\equiv
\gamma^{-1/3}(K_{ij}-K\gamma_{ij}/3)$, and an auxiliary three-vector,
$F_i\equiv\delta^{jk}\partial_j\tilde{\gamma}_{ik}$. Here
$\gamma_{ij}$ is the three metric, $K_{ij}$ the extrinsic curvature,
$\gamma\equiv\text{det}(\gamma_{ij})$, and $K\equiv
K_{ij}\gamma^{ij}$.  As in \cite{Shibata:2007zm}, we evolve the
conformal factor $\varphi$, not the inverse of $\psi$, because the
cell-centered grid is adopted in our code, and hence, the black hole
spacetime is handled in the moving puncture framework 
\cite{Campanelli:2005dd,bcckvm06}. 
For the conditions on the lapse, $\alpha$, and the shift vector, $\beta^i$, 
we adopt a dynamical gauge condition as in \cite{Shibata:2007zm}. 

The numerical scheme for solving the Einstein equation is the same as that 
in \cite{Shibata:2007zm}: We use the fourth-order finite difference 
scheme in the spatial direction and a third-order Runge-Kutta scheme in 
the time integration, where the advection terms such as 
$\beta^i\partial_i\varphi$ are evaluated by a fourth-order non-centered 
difference. 

The hydrodynamics equations are solved as in 
\cite{Shibata:2007zm}: We evolve 
$\rho_* \equiv \rho \alpha u^t e^{6\varphi}$, $\hat u_i
\equiv h u_i$, and $e_* \equiv \rho \alpha u^t -P/(\rho \alpha u^t)$,
where $\rho$ is the rest-mass density, $u_i$ is the three-component of
the four velocity, $P$ is the pressure, and $h$ is the specific
enthalpy defined by $h \equiv 1 + \varepsilon + P/\rho$ and
$\varepsilon$ is the specific internal energy defined by $\varepsilon \equiv \epsilon/\rho
- 1$. To handle advection
terms in the hydrodynamic equations, a high-resolution central scheme
\cite{KT} is adopted with a third-order piecewise parabolic
interpolation and with a steep min-mod limiter. In the present work,
the limiter parameter, $b$, is set to be $2.5$ (see 
\cite{Shibata2003} for detail about the parameter $b$).

The fluid evolution during inspiral is essentially free of shocks, and
when there are no shocks the simulations use the cold EOS specified in
Sec.~\ref{sec:modeleos}.  During merger, when the evolution has
shocks, we include a hot component with a thermal effective adiabatic
index $\Gamma_{\text{i}}$, as described in \cite{Shibata2003}.  Shock heating
in merger can increase the thermal energy up to $\sim 20$--30\% of
the total energy \cite{Shibata:2005ss}.

Gravitational radiation is extracted both by spatially decomposing the
metric perturbation about flat spacetime in the wave-zone with spin-2
weighted spherical harmonics and by calculating the outgoing part of the
Weyl scalar $\Psi_4$.  For equal mass neutron stars (such as those
studied here), the quadrupole $(\ell=2,m=\pm2)$ mode is much larger
than any other mode; we consider just this mode in this analysis.

The waveforms output from the simulations are the cross and plus
amplitudes $h_{+} c^2 D/GM_{\text{tot}}$ and $h_{\times} c^2
D/GM_{\text{tot}}$ of the quadrupole waveform, as would be measured at
large distance $D \gg GM_{\text{tot}}/c^2$ along the $z$ axis
perpendicular to the plane of the orbit, versus the retarded time
$t_{\text{ret}}$. Here $M_{\text{tot}}$ is the sum of the two neutron star
masses when they are far apart, $M_{\text{tot}}=2.7 M_\odot$. The
strain is sampled at discrete values evenly spaced in $t$, with a
sampling rate $\Delta t$ of between 0.006~ms (for 2B) and 0.031~ms
(for 2H) which depends on the time of simulation.  There is junk
radiation in the early part of the extracted waveforms for 
$t_{\text{ret}} \alt 0$. We discard this part in the data analysis. 

Each simulation is typically performed for three grid resolutions.  In the
best-resolution case, the diameter of neutron stars is covered by 60
grid points. Convergence tests with different grid resolutions
indicate that, with the best grid resolution, the time duration in the
inspiral phase is underestimated by about 1 orbit. This is primarily
due to the fact that angular momentum is spuriously lost by numerical
dissipation. Thus, the inspiral gravitational waves include a phase
error, and as a result, the amplitude of the spectrum for the inspiral
phase is slightly underestimated. However, we find that the waveforms
and resulting power spectrum for the late inspiral and merger phases, which
we are most interested in for the present work, depend weakly on the grid
resolution. 

\section{Waveform analysis}

We construct the complex quantity 
\begin{equation}h = h_{+} - i h_{\times}
\end{equation}
from the quadrupole waveform data.
The amplitude and phase of this quantity define the instantaneous amplitude
$|h|$ and phase $\phi = \arg h$ of the waveform. The
instantaneous frequency $f$ of the quadrupole waveform is then estimated by
\begin{equation}
f = \frac{1}{2 \pi} \frac{\Delta \phi}{\Delta t}
\end{equation}
The numerical data can be shifted in phase and time by adding a time shift
$\tau$ to the time series points and multiplying the complex $h$ by $e^{i
\phi}$ to shift the overall wave phase by $\phi$.

It is useful to define a reference time marking the end of the inspiral and
onset of merger. A natural choice for the end of the inspiral portion, 
considering the behaviour of a PP inspiral waveform, is the time of the peak
in the waveform amplitude $|h|$.
However, the amplitude of the numerical waveforms oscillates over the
course of an orbit. 
% (presumably due to orbital
%eccentricity, although extraction radius deformation could also
%contribute). 
A moving average of the waveform amplitude over $0.5$\,ms segments is used
to average this oscillation; the end of the inspiral is then defined as the
time at the end of the maximum amplitude interval. The resulting merger
time $t_{\text{M}}$ will be marked by solid vertical lines in the plots to
follow.

The numerical waveforms begin at different orbital frequencies. To align
them for comparison, they are each matched in the early inspiral region to
the same post-Newtonian point-particle (or PP) waveform.

\begin{figure}[!htb]
\caption[Numerical waveforms aligned by PN match]{Solid lines show
numerical waveforms, scaled by
$c^2D/GM_{\text{tot}}$, and aligned in time and phase to the same
point-particle post-Newtonian inspiral (dashed line), using the method
described in Sec.~\ref{sec:pnmatching}. The two dashed vertical bars
indicate the portion of the waveform used for matching; the last vertical
bar indicates the end of inspiral time $t_{\text{M}}$ for the numerical
waveform.  The top four simulations, 2H, H, HB, and B, show the start of
post-merger oscillations from a hypermassive NS remnant in the simulation.
2B shows quasinormal ringdown from a prompt collapse to a black hole
following merger.
\label{fig:matchalign}
}
\begin{center}
\includegraphics[width=85mm]{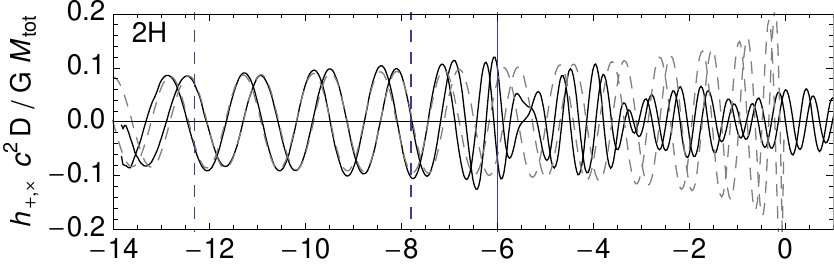} \\
\includegraphics[width=85mm]{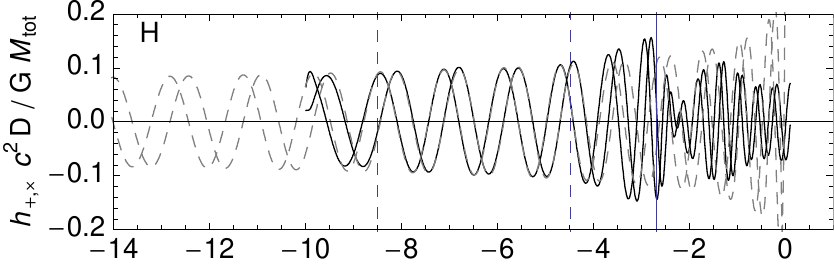} \\
\includegraphics[width=85mm]{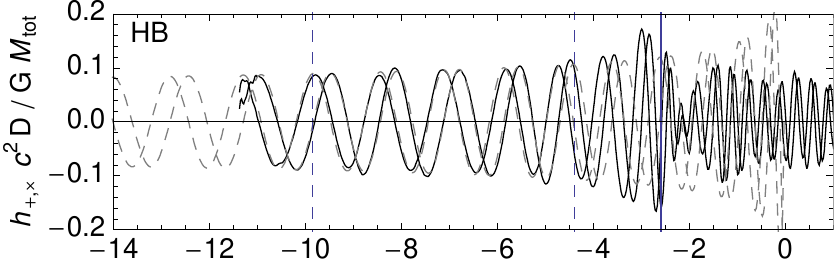} \\
\includegraphics[width=85mm]{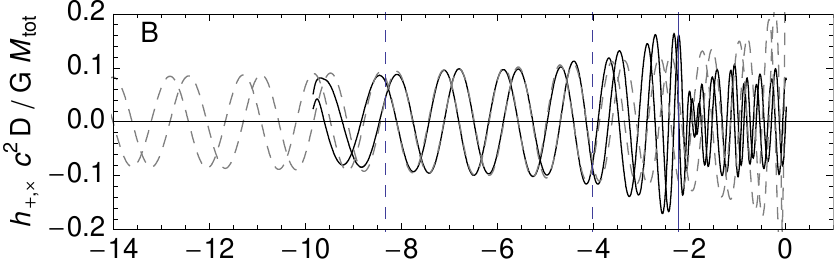} \\
\includegraphics[width=85mm]{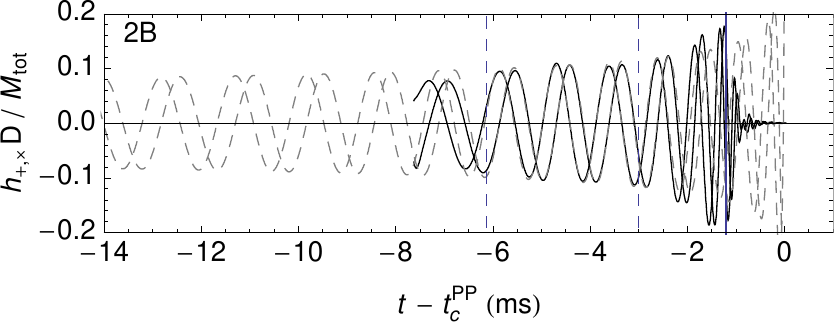}
\end{center}
\end{figure}

\subsection{Post-Newtonian point particle}

In full GR, point-particle inspiral is not well-defined, and one is left
with the post-Newtonian point-particle (PP) approximation and fully general
relativistic black-hole numerical solutions as natural substitutes.
Fortunately, the Taylor T4 3.0/3.5 post-Newtonian specification, introduced in
\cite{Boyle:2007}, agrees closely with numerical binary black hole
waveforms for many cycles, up to and including the cycle before merger (see
also \cite{Gopakumar:2007}).  This empirical agreement allows us to adopt
the Taylor T4 waveform as an appropriate PP baseline waveform, compatible
with full GR until the last cycles. We will show that the binary
neutron star waveforms depart from this waveform $4$--$8$ cycles (200--560
$M_{\text{tot}}$) before the best-fit PP merger.

The TaylorT4 waveform is constructed by numerically integrating for a
post-Newtonian parameter $x$, which is related to the orbital frequency
observed at infinity $\Omega$: 
\begin{equation}
\label{eq:xdef}
x = \left( \frac{ G M_{\text{tot}} \Omega }{ c^3 } \right)^{2/3},
\end{equation}
where $M_{\text{tot}}$ is the sum of the point masses.
To first post-Newtonian order, $x \sim M_{\text{tot}}/r$ where $r$ is the
orbital radius.

Following \cite{Boyle:2007}, we use
\begin{equation}
\begin{split}
\label{eq:dxdt1}
\frac{dx}{dt} =& \frac{16}{5} \frac{ c^3}{G} \frac{x^5 }{M_{\text{tot}}}
\left[
  1 - \frac{487}{168} x + 4 \pi x^{3/2} 
  + \frac{274\,229}{72\,576} x^2
\right.
\\
&\left.
  - \frac{254}{21} \pi x^{5/2}
  + \left\{
    \frac{178\,384\,023\,737}{33\,530\,011\,200} + \frac{1475}{192}\pi^2
	\right.
\right.
\\
&\left.
		\left.
    - \frac{1712}{105}\gamma
  	- \frac{856}{105}\ln(16x)
  \right\} x^3
  + \frac{3\,310}{189} \pi x^{7/2} 
\right] ,
\end{split}
\end{equation}
to find $x(t)$ and then orbital phase of the binary is found from
\begin{equation}
\label{eq:dphidt}
\frac{d\Phi}{dt} = \frac{c^3}{G M_{\text{tot}}} x^{3/2}.
\end{equation}
This yields a result to 3.5 post-Newtonian order for the phase evolution.
The constants of integration are fixed by specifying the coalescence time
$t_{\text{c}}^{\text{PP}}$, when $x \to \infty$, and the orbital phase at
this time $\Phi(t_{\text{c}}^{\text{PP}})=\Phi_{\text{c}}^{\text{PP}}$,

The amplitude of
the quadrupole waveform is calculated to 3.0 order in 
\cite{Kidder:2008}. For the complex waveform $h_{\text{PP}}$ measured
along the orbital axis of the binary, using the
appropriate spherical harmonic conventions, the result is
\begin{equation}
\label{eq:h3pn}
\begin{split}
h_{\text{PP}} =& - \frac{G}{c^2} \frac{M_{\text{tot}}}{D} e^{2 i \Phi}
x \left[
  1- \frac{373}{168} x
  + 2\pi x^{3/2}
  - \frac{ 62\,653}{24\,192}x^2
\right.\\
&\left.
 - \left\{
    \frac{197}{42}\pi + 6 i \right\}x^{5/2}
  + \left\{ 
    \frac{43\,876\,092\,677}{1\,117\,670\,400} 
		+ \frac{99}{128}\pi^2 
  \right.
\right.\\
&\left. 
  \left.
	 -\frac{856}{105}\gamma 
  - \frac{1712}{105}\ln(2) 
    - \frac{428}{105}\ln(x)
    + \frac{428}{105} i \pi
  \right\} x^3
\right]
\end{split}
\end{equation}
where $D$ is the distance to the observer, and logarithmic terms with a
frequency scale have been absorbed into the the phase.  The coalescence
phase of the quadrupole point particle waveform,
$\phi_{\text{c}}^{\text{PP}}=\arg h_{\text{PP}}(t_{\text{c}}^{\text{PP}})$,
is determined by a choice of the orbital coalescence
phase $\Phi_{\text{c}}^{\text{PP}}$.

\subsection{Match to post-Newtonian point particle}
\label{sec:pnmatching}
To match the numerical data to the PP inspiral waveform, a relative time shift
and a relative phase shift must be specified by varying these parameters to
obtain the best match.  The masses of the point particles in the PP
waveform are fixed to be the same as the neutron stars in the numerical
simulations -- the gravitational mass of isolated (TOV) neutron stars
with the same number of baryons -- and so masses are not varied in finding
the best match.  With a goal of signal analysis, we choose the time and
phase shift by maximizing a correlation-based match between two waveforms.

The complex numerical relativity derived quadrupole waveform,
$h_{\text{NR}} = h_+^{\text{NR}} - i h_\times^{\text{NR}}$,
is convolved with the complex post-Newtonian quadrupole waveform,
$h_{\text{PP}} = h_+^{\text{PP}} - i h_\times^{\text{PP}}$ of
Eq.~(\ref{eq:h3pn}), over a given matching region $T_{\text{I}}<t<T_{\text{F}}$.
\begin{equation}
z(\tau;T_{\text{I}},T_{\text{F}}) = \int_{T_{\text{I}}}^{T_{\text{F}}} h_{\text{NR}}^{\vphantom{\ast}}(t) h_{\text{PP}}^{\ast}(t - \tau) dt.
%    &=& \int_{T_{\text{I}}}^{T_{\text{F}}} \left[h_+(t) - i h_\times(t)\right] 
% \left[h^{\text{PP}}_+(t-\tau) + i h^{\text{PP}}_\times(t-\tau)\right] dt\\
    %&=&  \int_{T_{\text{I}}}^{T_{\text{F}}} \left[h_+(t) h^{\text{PP}}_+(t-\tau)  
            %+ h_\times(t) h^{\text{PP}}_\times(t-\tau)\right] dt \nonumber\\
    %&+&\, i \int _{T_{\text{I}}}^{T_{\text{F}}} \left[ h_+(t) h^{\text{PP}}_\times(t-\tau)  
            %- h_\times(t) h^{\text{PP}}_+(t-\tau)\right] dt
\end{equation}
This quantity is similar to the complex matched filter output of the
\textsc{findchirp} algorithm \cite{findchirp}. The real part of
$z(\tau;T_{\text{I}},T_{\text{F}})$ corresponds to the correlation of the two waveforms as a
function of the time shift $\tau$.  The absolute value $|z(\tau;T_{\text{I}},T_{\text{F}})|$ is
the correlation maximized over a constant overall phase shift, and the argument
is the overall phase shift required to so maximize.  A normalized match between
two waveforms is
\begin{equation}
  m(\tau; T_{\text{I}}, T_{\text{F}}) = \frac{z(\tau;T_{\text{I}},T_{\text{F}})}%
  {\sigma_{\text{NR}}(0;T_{\text{I}},T_{\text{F}})\,\sigma_{\text{PP}}(\tau;T_{\text{I}},T_{\text{F}})}
\end{equation}
where the normalizing constants are defined by
\begin{equation}
  \sigma^2(\tau;T_{\text{I}},T_{\text{F}}) = \int_{T_{\text{I}}}^{T_{\text{F}}} |h(t-\tau)|^2 \, dt
\end{equation}
where $h$ is either $h_{\text{NR}}$,
which defines $\sigma_{\text{NR}}(\tau;T_{\text{I}},T_{\text{F}})$, or $h_{\text{PP}}$,
which defines $\sigma_{\text{PP}}(\tau;T_{\text{I}},T_{\text{F}})$.

\begin{figure}
\caption{Phase-optimized, time-limited match between numerical inspiral
waveforms and point particle post-Newtonian waveforms. Contours are shown
at 0.95, 0.98, 0.99, and 0.997. as a function of match region truncation at
some time before numerical merger, $T_{\text{F}}-t_{\text{M}}$, and relative shift between
numerical and point particle waveforms, $t_{\text{c}}^{\text{PP}} - t_{\text{M}}$. The start of the
match is fixed to 1.5 ms after the start of the numerical waveform.
Subsequent analysis in this paper is done using the best match at a fixed
$T_{\text{F}}-t_{\text{M}}$ of 1.8 ms for each waveform.
\label{fig:taut}}
\begin{center}
\includegraphics[height=53mm]{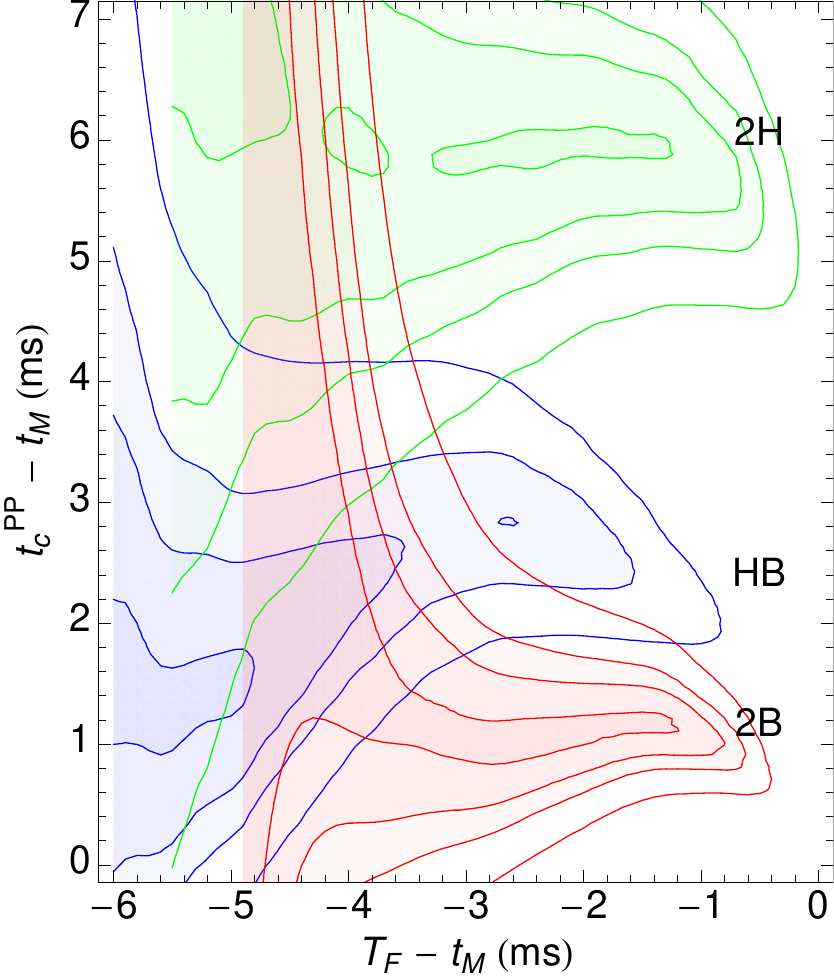} 
\includegraphics[height=53mm]{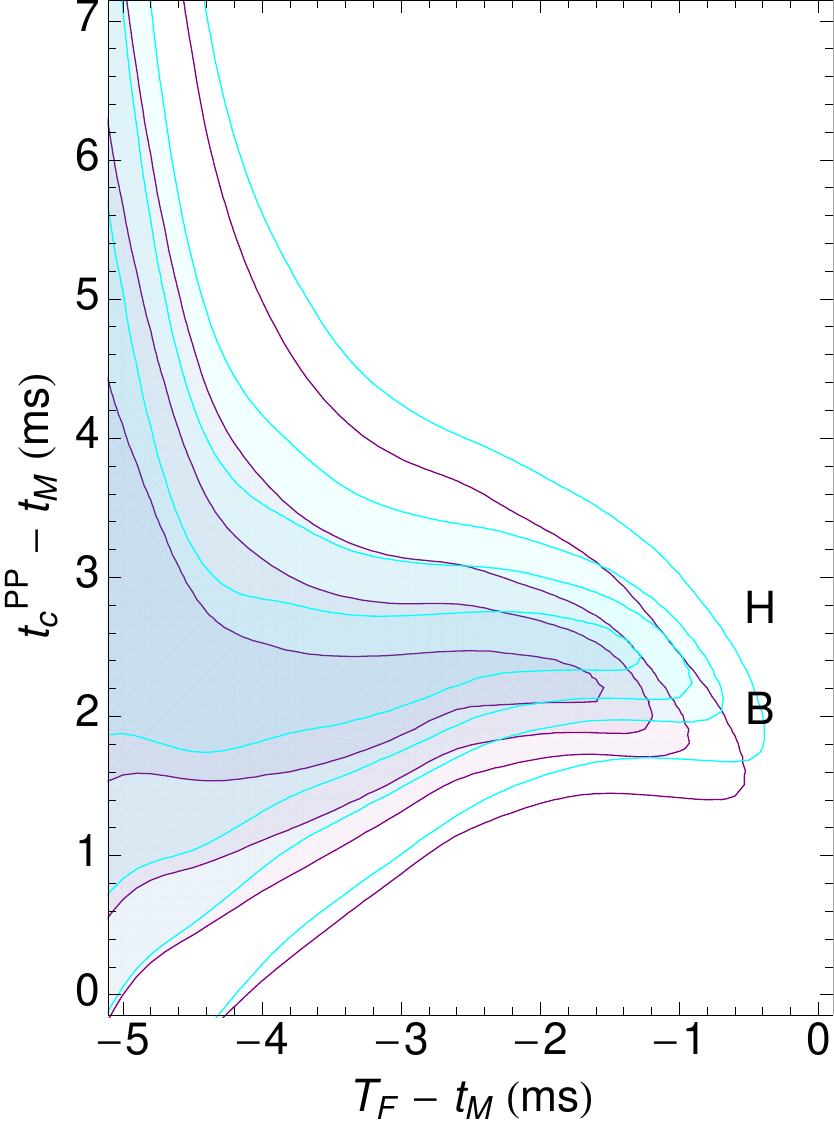} 
\end{center}
\end{figure}

The shift $\tau$ maximizing the match between two waveforms is sensitive
to the portion of the numerical waveform matched, $T_{\text{I}}<t<T_{\text{F}}$. Some
truncation of the numerical waveform is required to eliminate residual
effects of initial data. One would also like to truncate the waveform
at some point before the peak amplitude, at the end of the region without
significant finite size effects. To determine where this region is,
we plot the match between numerical waveforms as a function of both the end
point of the match $T_{\text{F}}$, and the shift $\tau$, measuring quantities 
relative to the previously defined merger time $t_{\text{M}}$ of each numerical
waveform and the PP coalescence time $t_{\text{c}}^{\text{PP}}$ so that $\tau =
t_{\text{c}}^{\text{PP}} - t_{\text{M}}$, in Fig.~\ref{fig:taut}.

Reassuringly, for most of the waveforms, once we truncate a millisecond or
so before the numerical merger $t_{\text{M}}$, there is a region where the
range of well-matched shifts $\tau$ show little dependence on the exact end
point $T_{\text{F}}$. This continues with earlier end points until the
segment of the waveform considered becomes so short that the range of
matching $\tau$ broadens significantly. The exception is the waveform of
HB, which was the first simulation of the series. HB exhibits somewhat
larger eccentricity, and drift in the best match timeshift, compared to the
other waveforms.

For subsequent analysis, we choose to take $\tau$ maximizing the match for a
region of the numerical waveforms between $1.5$\,ms after the start of the
waveform and $1.8$\,ms before numerical merger.  Varying the details of
this choice can change the best match $\tau$ by up to $\simeq 1$\,ms. By
comparison, one waveform cycle takes between $0.5$ and $2$\,ms in the
inspiral region, so this will be a significant source of uncertainty in SNR
estimates.  Longer or more accurate simulations are required to more
precisely fix a post-Newtonian match.

\subsection{Comparison of waveforms}

%
%\begin{figure}[!htb]
%\caption{ Time-amplitude behaviour, vertical line
%markings as previous
%figure.
%\label{fig:timeamp}}
%\begin{center}
%\begin{tabular}{cc}
%\includegraphics[width=60mm]{timeamp2B.pdf} \\
%\includegraphics[width=60mm]{timeamp5B.pdf} \\
%\includegraphics[width=60mm]{timeamp1B.pdf} \\
%\includegraphics[width=60mm]{timeamp4B.pdf} \\
%\includegraphics[width=60mm]{timeamp3B.pdf}
%\end{tabular}
%\end{center}
%\end{figure}

Unlike the case of matching binary black hole simulations to point particle
post-Newtonian\cite{Boyle:2007,Gopakumar:2007}, the binary neutron star simulations show departure
from point particle many cycles before the post-Newtonian merger time.
Fig.~\ref{fig:matchalign} shows the four waveforms shifted so the
best-match PP waveforms have the same $t_{\text{c}}^{\text{PP}}$ and
$\phi_{\text{c}}^{\text{PP}}$. As the stiffness of the EOS and thus
the radius of the neutron stars, increases, the end of inspiral for the
binary neutron stars is shifted away from the end of inspiral for point
particle post-Newtonian.

\begin{figure}[!htb]
\caption{Time-frequency behavior, vertical line
markings as previous figure. The departure from the point particle
time-frequency relations, shown using a long-dashed line, occurs between
$700$--$1000$\,Hz depending on EOS.
 \label{fig:timefreq}}
\begin{center}
\begin{tabular}{cc}
\includegraphics[width=60mm]{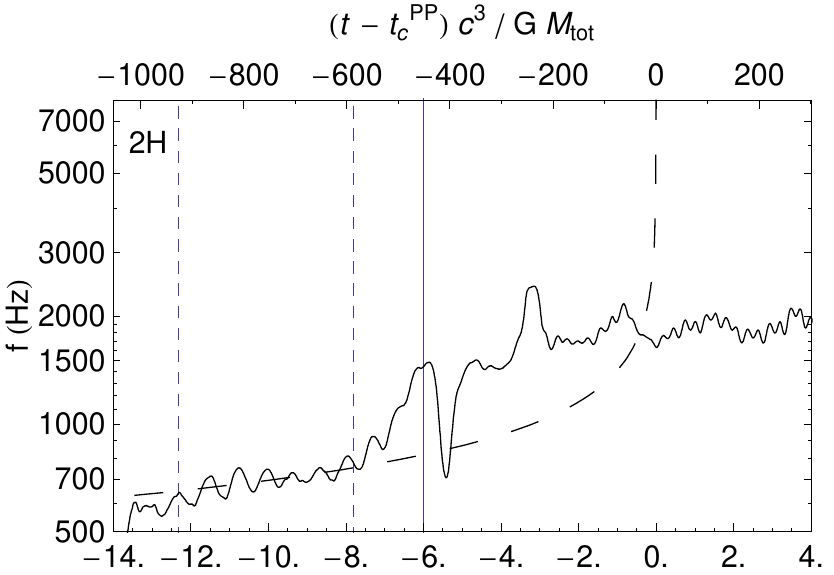} \\
\includegraphics[width=60mm]{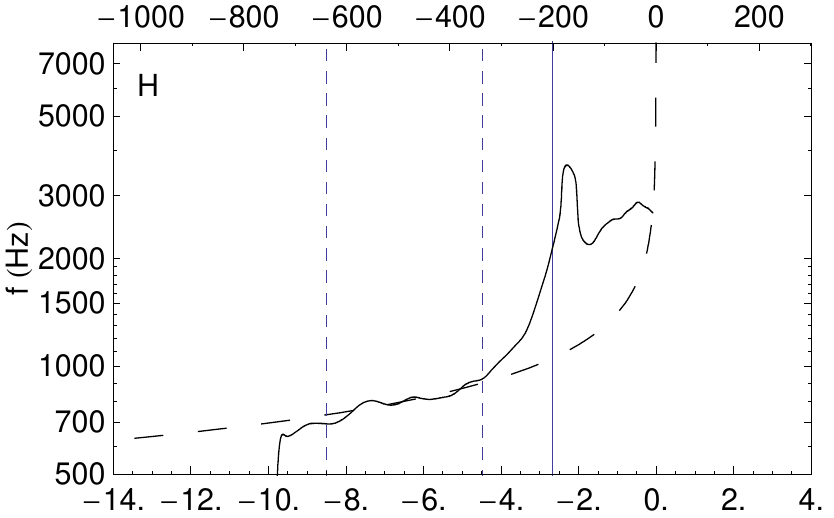} \\
\includegraphics[width=60mm]{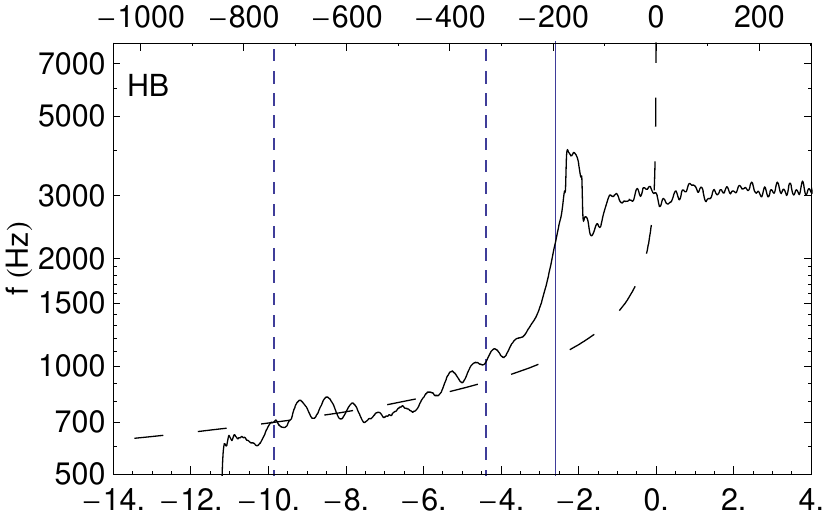} \\
\includegraphics[width=60mm]{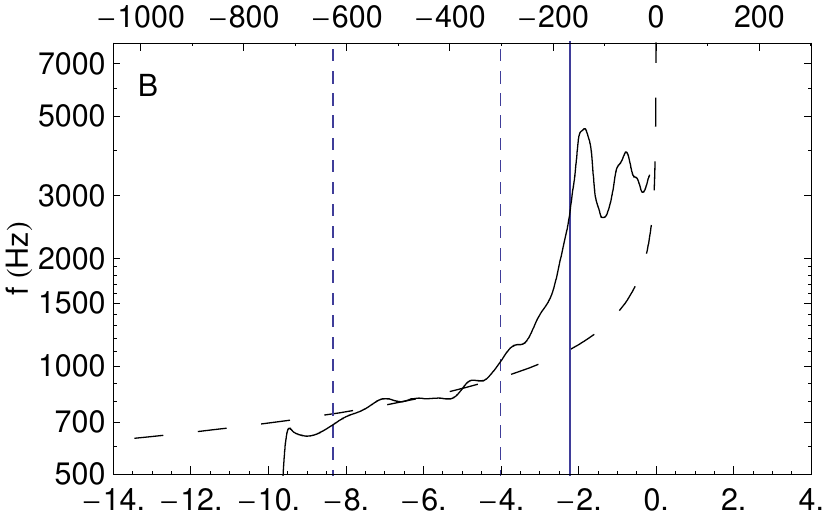} \\
\includegraphics[width=60mm]{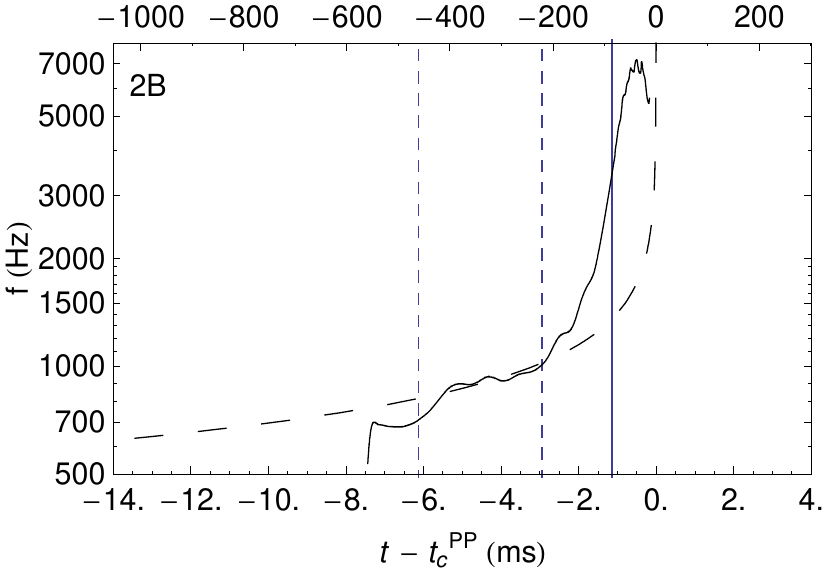}
\end{tabular}
\end{center}
\end{figure}

This can also be seen by
plotting the instantaneous frequency of the numerical simulation waveform
with the same time shifts, as in Fig.~\ref{fig:timefreq}, which also
shows more clearly the difference in the post-merger oscillation
frequencies of the hyper-massive remnants, when present. The larger neutron
star produced by the stiff EOS 2H has a lower oscillation frequency than
that from the medium EOS HB. 
%[[Are they all hypermassive? - check 2H max
%mass, it is super-stiff so maximum mass may be very large, check amount of
%matter thrown off realistic eos runs.]] 
The remnant forms with a bar-mode
oscillation stable over a longer period than the $\sim10$\,ms % [[refine]]
simulated. The signal from such a bar mode may be even stronger when the
full lifetime is included, although aspects of the physics neglected in
this study will likely come into play. Information that can be extracted
from the presence (or absence) and characteristics of a post-merger
oscillation signal would complement the information present in the late
inspiral. This is another subject for future study.

\subsection{Frequency spectrum of waveforms}
Given $h_+$ or $h_\times$, one can construct the discrete Fourier
transforms (DFTs) $\tilde{h}_+$ or $\tilde{h}_\times$. Both polarizations
yield the same DFT amplitude spectrum $|\tilde{h}|$, with phase shifted by
$\pi/2$, if one neglects discretization, windowing, and numerical effects
(including eccentricity).  The amplitude spectrum $|\tilde{h}|$ is
independent of phase and time shifts of the waveform.
%[[[double check this]]

The stationary phase approximation is valid for the post-Newtonian waveform
up to frequencies of about 1500\,Hz (with $\lesssim 10\%$ error), so is used to
plot the amplitude of the point particle spectrum. In terms of an
instantaneous frequency
\begin{equation}f(t) = \frac{1}{2 \pi} \frac{d\phi}{dt},\end{equation}
the Fourier transform of the waveform has the amplitude
\begin{equation}
|\tilde{h}| \simeq A(f) \left(\frac{df}{dt}\right)^{-1/2} .
%\mathrm{exp}\left(i\left(2\pi
%f t_{\text{c}} - \phi(f) - \frac{\pi}{4}\right)\right).
\end{equation}
For binaries comprised of equal mass companions, the gravitational radiation
is dominated by quadrupole modes throughout the inspiral.  The wave phase
$\phi$ is negligibly different from twice the orbital phase $2\Phi$ until
the onset of merger at very high frequencies so we can use the relation
\begin{equation}
\frac{df}{dt} \simeq \frac{1}{\pi}\frac{c^3}{G M} \frac{3}{2} x^{1/2}
\frac{dx}{dt};
\end{equation}
to write the amplitude of the Fourier transform entirely in
terms of the functions $dx/dt$ of Eq.~(\ref{eq:dxdt1}) and the amplitude
of the polarization waveforms $A(f) = |h|$ in
Eq.~(\ref{eq:h3pn}).

\begin{figure*}
\caption[DFT of full numerical waveforms]{DFT of full numerical waveforms,
at an effective distance $D_{\text{eff}}=100$\,Mpc, compared to
noise spectra for Advanced LIGO (labelled ``AdvLIGO'' for the standard
configuration and ``Broadband'' for the broad-band configuration) and the
Einstein Telescope (labelled ``ET'') shown by thick grey lines. The DFT of the
numerical waveforms turned off after $t_{\text{M}}$ is shown by dot-dashed lines, the stationary-phase point particle is shown by a dashed line for
reference. The lower right figure shows a combined plot of
inspiral-truncated waveforms, smoothly joined on to best-match PP inspiral
time series before the DFT is taken. 
\label{fig:dft}}
\begin{tabular}{ll}
\includegraphics[width=70mm]{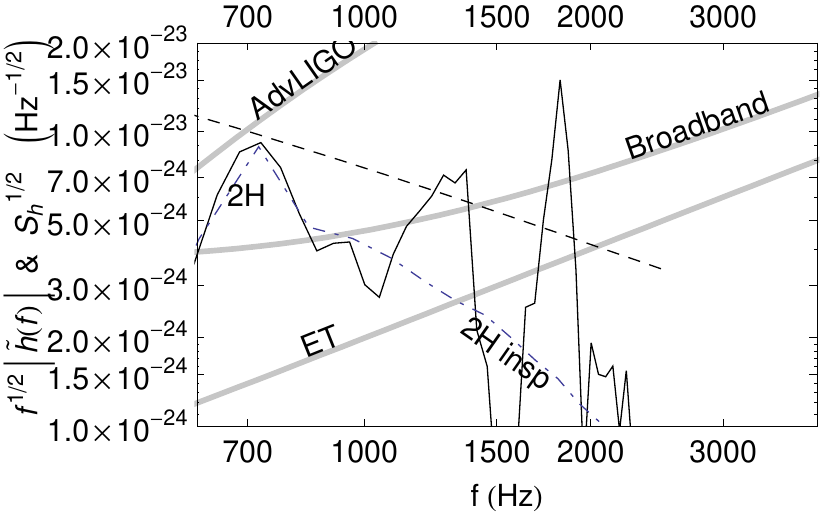} 
\includegraphics[width=70mm]{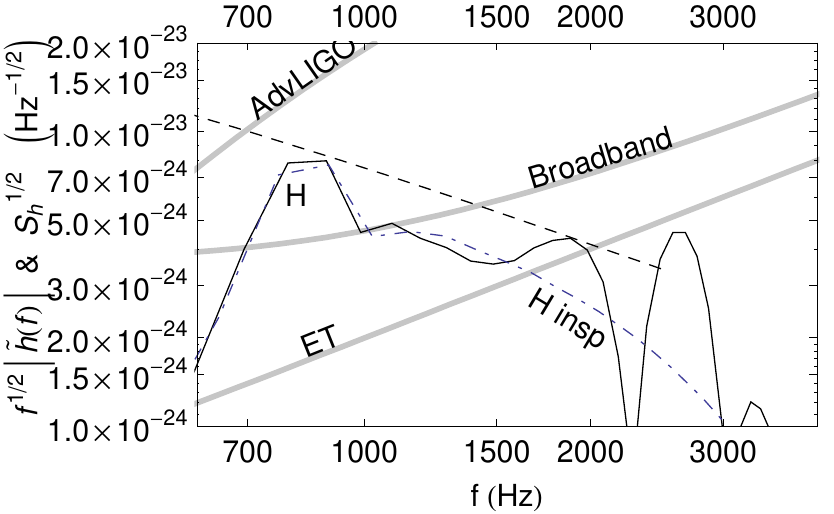}\\ 
\includegraphics[width=70mm]{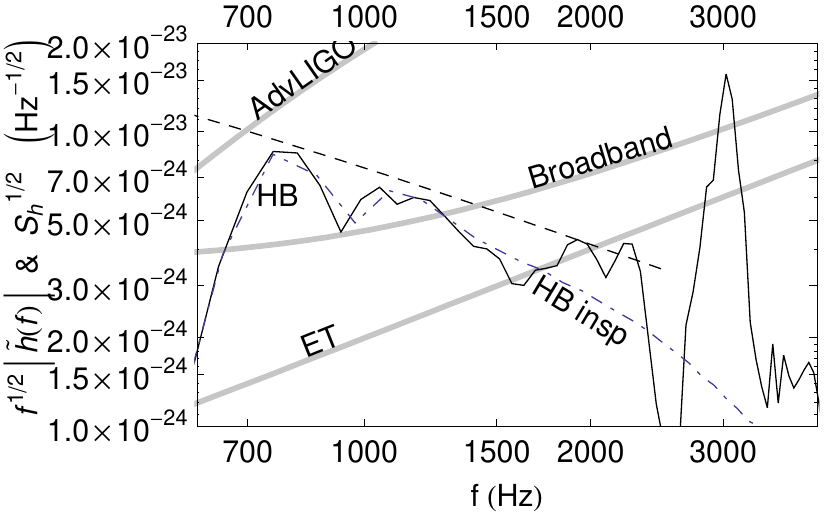}
\includegraphics[width=70mm]{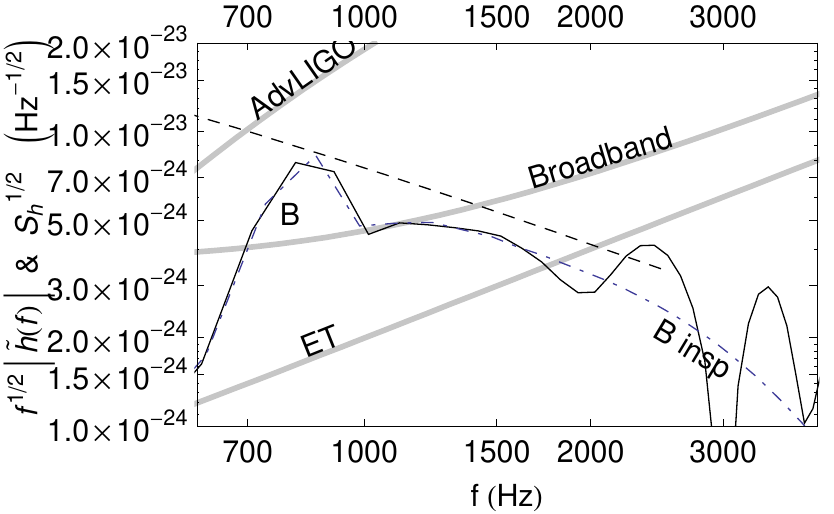}\\ 
\includegraphics[width=70mm]{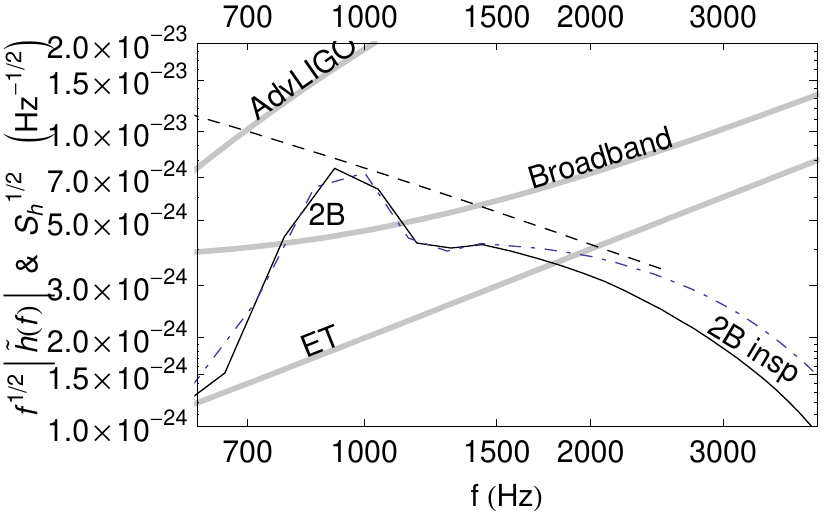}
\includegraphics[width=70mm]{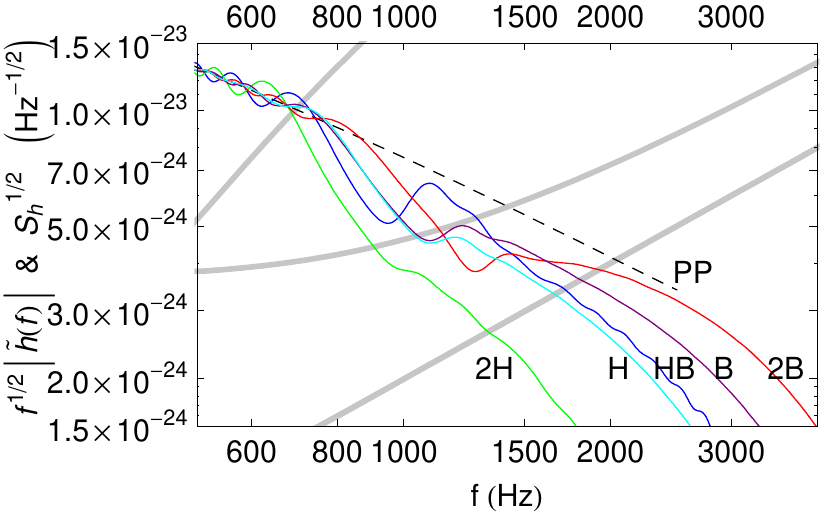} \\
\end{tabular}
\end{figure*}
%\begin{figure}[!htb]

The translation of emitted waveforms into the strain amplitude measured at
a detector involves transformations incorporating the effects of the
emitting binary's angle of inclination and sky location. 
%[[ For a quadrupole
%waveform, such as those considered here, the plus and cross polarizations
%at the detector are straightforwardly related somethiing something maybe]] 
These effects are absorbed into an effective distance $D_{\text{eff}}$,
which is equal to the actual distance $D$ for a binary with  optimal
orientation and sky location, and is greater than the actual distance for
a system that is not optimally oriented or located.
The detector will detect a single polarization of the waveform, some
combination of the plus and cross polarizations of the emitted waveform.
The polarizations extracted from the simulation can be used as two
estimates of the strain at the detector for a given numerically modelled
source, which give very close results and are subsequently averaged.

To compare with noise curves in the usual units, the quantity $f^{1/2}
|\tilde{h}(f)|$ is plotted, at a reference distance of
$D_{\text{eff}}=100$\,Mpc and rescaling from previously plotted numerical
output $h(t) c^2D/GM_{\text{tot}}$ using $M_{\text{tot}} = 2.7 M_\odot$.

The full spectra of models 2H, H, HB, and B, seen in Fig. \ref{fig:dft},
show peaks at post-merger oscillation frequencies; those of H and B are
weaker as the waveform is truncated shortly after the formation of the
hyper-massive remnant. Waveforms of 2H and HB are also truncated while the
post-merger oscillation is ongoing; if the simulations were allowed to
continue, these peaks would presumably grow further. The simulation of 2B,
in contrast, collapses to a black hole and has only a short lived (and
relatively small-amplitude) quasinormal mode ringdown post merger.

Note that time-frequency plots like the ones shown Fig.~\ref{fig:timefreq}
showed that numerical waveforms follow the PP waveform at instantaneous
frequencies of up to $700$--$1000$\,Hz, depending on the EOS\@. The
disagreement in the spectra in Fig.~\ref{fig:dft} from the PP
stationary-phase approximation waveform at frequencies below this is
primarily due to the finite starting time of the numerical waveforms.  To
estimate spectra from the full inspiral, we construct hybrid waveforms.
The short-term numerical waveforms are are smoothly merged on to
long-inspiral PP-PN waveforms with Hann windowing to smoothly turn on 
\begin{equation}
w(n) = \frac{1}{2}\left[1-\cos\left(\frac{\pi
n}{N-1}\right)\right]
\end{equation}
or turn off
\begin{equation}
w(n) = \frac{1}{2}\left[1+\cos\left(\frac{\pi n}{N-1}\right)\right]
\end{equation}
a signal over a range of $N$ points, $0\le n<N$ (or over a time $N\,\Delta t$).
To construct hybrid waveforms, these windows are used to turn on the
numerical waveform as the matched post-Newtonian is turned off, such that
the sum of the two window functions is 1 over the
matching range. 
%[[choice of matching range]]

Although the post-merger oscillations
are an interesting source of potentially measurable strains, the
dependence on the cold EOS is less straightforward as
temperature effects become significant during the the merger. We focus
instead on the signal from the waveforms during the inspiral region only,
turning off the waveforms after the end of inspiral, $t_{\text{M}}$. The
DFT amplitudes of inspiral-only hybrid waveforms are also plotted in
Fig.~\ref{fig:dft} from the range of EOS considered. We see here that HB
is estimated to depart from PP earlier than either H or B, rather than at
the expected intermediate value. This is ascribed to the higher
eccentricity and lower accuracy of the match for the HB waveform.

\section{Detectability}
The question for neutron star astrophysics is whether these
differences in the waveform will be  measurable. We consider the
possibility of detecting EOS effects with Advanced LIGO style detectors
in varying configurations.

We use several detector configurations commonly considered for Advanced LIGO
include tunings optimized for 1.4~$M_\odot$ NS-NS inspiral detection
(``Standard''), for burst detection (``Broadband''), and for pulsars at
1150\,Hz (``Narrowband'').  The sensitivity is expressed in terms of the
one-sided strain-equivalent amplitude spectral density $S_h(f)$ (which has
units of $\mbox{Hz}^{-1/2}$) of the instrumental noise in Advanced LIGO\@.
We also consider a provisional noise curve for the Einstein Telescope
\cite{ETnoise}. We consider only a single detector of each type, rather
than a combination of detectors, for a preliminary estimate of
detectability.
 
Given two signals $h_1$ and $h_2$, define the usual inner product for a
given noise spectrum $S_h(f)$ \cite{CutlerFlanagan1994}:
\begin{equation}
\label{eq:sigmetr}
\left( h_1 | h_2 \right) = 4 \mathop{\text{Re}}\nolimits\int_{0}^{\infty} \frac{
\tilde{h}_1(f) \tilde{h}_2^*(f) }{S_h(f)} df .
\end{equation}
This inner product yields a natural metric on a space of waveforms with
distance between waveforms weighted by the inverse of the noise. We filter
the detector output $s$ against an expected waveform $h$ using $\left( s | h
\right)$. Then
\begin{equation}
\varrho =\frac{ \left( s | h \right)}{\sqrt{\left( h | h
\right)}}
\end{equation}
is the optimal statistic to detect a waveform of known form $h$ in the
signal $s$.  If the detector output contains a particular waveform $h$
that is exactly matched by the template used, then the expectation value
of $\varrho$ is the expected signal-to-noise ratio or SNR of that waveform:
\begin{equation}
\overline{\varrho} = \sqrt{\left( h | h \right)}.
\end{equation}

Given a signal $h_1$ to be used as a template, one
can consider whether a known \emph{departure} from this signal can be
measured.  Assuming the modified waveform $h_2$ is known, the expected
SNR of $h_2 - h_1$ is similarly
\begin{equation}
\overline{\varrho_{\text{diff}}} = \sqrt{\left( h_2 - h_1 | h_2 -
h_1 \right)}.
\end{equation}
The two signals will be (marginally) distinguishable\footnote{Compare
discussion in \cite{LindblomOwenBrown} of
indistinguishability.}.  if the difference between the waveforms has
$\overline{\varrho_{\text{diff}}} \geq
1$.

We analyse the measurability of the differences between hybrid
inspiral-only waveforms matched to point particle PN waveforms in the early
inspiral.  After the numerical waveforms have been matched to the same
post-Newtonian point-particle inspiral, the signals will be aligned in time
and phase.  We can then then compare the resulting waveforms to each other,
and to a PN-only waveform, using different Advanced LIGO noise spectra.

We report results in terms of the SNR measured by a single
detector at an effective distance $D_{\text{eff}}=100$\,Mpc. Results can
be scaled to any distance; as each measured $h \propto 1/D$,
$\overline{\varrho_{\text{diff}}}
\propto 1/D$.
%\begin{equation}
%\varrho_{\mathrm{diff},D} = \varrho_{\mathrm{diff},
%100\,\mathrm{Mpc}} \left( \frac{100\,\mathrm{Mpc}}{D} \right).
%\end{equation}

%\begin{table}[!htb]
\begin{table}[t]
\caption{$\overline{\varrho_{\text{diff}}}$ in standard (NS-NS detection optimized) noise $\times
\left( 100\,\mbox{Mpc} / D_{\text{eff}} \right)$ \label{tab:rdstd}}
\begin{center}
\begin{tabular}{c  c c c c c}\hline
Model &  2B & B & HB & H &2H\\
 \hline\hline
PP &  0.32 &	0.45 &	0.55 &	0.46 &	0.69 \\
2B &  0& 0.36 &	0.48 &	0.38 &	0.63\\
B  &  & 0& 0.21 &	0.12	 &0.58\\
HB &  & & 0& 0.27 & 0.60\\ 
H & &  & & 0 & 0.58\\
\hline
\end{tabular}
\end{center}
\end{table}

%\begin{table}[!htb]
\begin{table}[t]
\caption{$\overline{\varrho_{\text{diff}}}$ in broadband (burst-optimized) noise $\times \left(
100\,\mbox{Mpc} / D_{\text{eff}} \right)$ \label{tab:rdbb}}
\begin{center}
\begin{tabular}{c  c c c c c}\hline
Model  & 2B & B & HB & H&2H\\
\hline\hline
PP &1.86& 2.32& 2.67& 2.38& 2.89\\
2B & 0&1.92& 2.32& 2.03& 2.54\\
B  & & 0& 0.81& 0.80& 2.27\\
HB  & & & 0& 1.28& 2.37\\
H  & & & & 0&  2.35 \\
\hline
\end{tabular}
\end{center}
\end{table}
%\begin{table}[!htb]
\begin{table}[t]
\caption{$\overline{\varrho_{\text{diff}}}$ in narrowband $1150$\,Hz noise $\times \left(
100\,\mbox{Mpc} / D_{\text{eff}} \right)$ \label{tab:rdnb} }
\begin{center}
\begin{tabular}{c c c c c c c} \hline
Model &  2B & B & HB & H &2H\\ \hline\hline
PP &  0.91& 2.75& 3.69& 2.65& 2.12\\
2B & 0&1.92& 2.92& 1.82& 1.45\\
B & & 0& 1.14& 0.22& 1.43\\
HB &  & & 0& 1.34 & 2.25 \\
H &  & & & 0&  1.42 \\
\hline
\end{tabular} 

\end{center}
\end{table} 

The difference between waveforms due to finite size effects is not
detectable in the NS-NS detection optimized configuration of Advanced LIGO
for $\sim 100$\,Mpc effective distances. \emph{However, in both narrowband and
broadband the differences can be significant}, and waveforms are
distinguishable from each other and from the PP waveform. Note that this
implies that even if the details of a NS-NS late inspiral signal are not
known, the difference between it and a point particle waveform should be
measurable.  The quantity $\overline{\varrho_{\text{diff}}}$ between the
observed waveform and a best fit point particle waveform, limited to
differences at high frequency, may be useful in itself to constrain
possible EOS without reference to waveform details.

%Given an effective distance $D_eff$, the volume
%of space which is enclosed within this distance is [[blah]]. [[Something
%about multiple detectors.]]

\subsection{Parameter estimation}
We will assume that EOS effects on the waveform impact the late inspiral
only.  For simplicity, we assume that orbital parameters, such as
$M_{\text{tot}}$, mass ratio $\eta$, point particle  post-Newtonian
coalescence time $t_{\text{c}}^{\text{PP}}$, and phase shift
$\phi_{\text{c}}^{\text{PP}}$, are determined from the observations of the
earlier inspiral waveform, with
sufficient accuracy that their measurement uncertainty will not affect the
accuracy to which the late inspiral effects determine the EOS parameters.
These measurements would be made by a broad-band instrument, in which the
signal-to-noise ratio is expected to be high ($\sim 40$ at 100\,Mpc) and
measurement accuracy is expected to be good~\cite{CutlerFlanagan1994}.
Inaccuracies in these measurement could lead to biases in the measured
EOS\@.  This will be an important aspect to assess when
high-quality binary neutron star simulations with various masses become
abundant.

With a one-parameter family of waveforms sampled, we can estimate the
accuracy to which this parameter can be measured.
There are also other EOS-related parameters which are not considered. In
the direct analysis of the measurability of the EOS parameter $p_1$, we
ignore variations of $\Gamma$ within the core. Correspondingly, in the analysis of radius
measurement, variations of the internal structures are neglected. 
Expanding coverage of the EOS parameter space is an ongoing project.
However, these initial parameter choices are expected to give the dominant
contributions to finite size effects of the waveform%
\footnote{ For example, using the 1PN tidal effect estimates of
\cite{FlanaganHinderer2007,Hinderer2008} yields about 10\% variation in
the tidal terms contributing to binding energy and luminosity from changing
internal structure---varying the apsidal constant---while keeping radius
fixed.}.

We estimate errors in parameter
estimation to first order in
$1/\overline{\varrho}$ or, equivalently, in $\delta\theta^A$, using the 
Fisher matrix $\Gamma_{AB} = \left(
\partial_A h | \partial_B h\right)$ \cite{CutlerFlanagan1994}. Its inverse, $(\Gamma^{-1})^{AB}$,
yields
\begin{equation}
\overline{\delta \theta^A \delta \theta^B} = (\Gamma^{-1})^{AB}
\end{equation}
so that the expected error in a given parameter $\theta^A$ is
\begin{equation}
\overline{\left(\delta \theta^A \right)^2} = (\Gamma^{-1})^{AA}
\end{equation}
and the cross terms of the inverse Fisher matrix yield correlations between
different parameters.

With a few simulations of varying parameter value, we 
estimate  $\partial h/\partial {p_1}$ and $\partial h/\partial {R}$ from
two of the sampled waveforms, $h_1$ and $h_2$.  For a single parameter $\theta$
(which can be taken to be either $p_1$ or $R$), we have
\begin{equation}
\left. \frac{\partial h}{\partial {\theta} } \right|_{\theta=(\theta_{1}+\theta_{2})/2} 
\simeq \frac{ h_2 - h_1 }{\theta_{2} - \theta_{1}}
\end{equation}
where $h_1=h(\theta_1)$ and $h_2=h(\theta_{2})$,
and then, for our one-parameter family where we neglect correlations with other
parameters, we have to first order
\begin{equation}
\overline{\left(\delta \theta \right)^2} \simeq 
\frac{(\theta_{2} - \theta_{1})^2}
{\left( h_2 - h_1 | h_2 - h_1  \right)}.
\end{equation}

Using adjacent pairs of models to estimate waveform dependence at an average
parameter value, we then find estimates of radius measurability as shown in
Table~\ref{tab:deltar} and $p_1$ measurability as shown in
Table~\ref{tab:deltap} for the burst-optimized noise configuration.

%\begin{table}[!htb]
\begin{table}[t]
\caption{$\delta R$ (km) in broadband (burst-optimized) noise $\times \left(
D_{\text{eff}}/ 100\,\mbox{Mpc}  \right)$ \label{tab:deltar}}
\begin{center}
\begin{tabular}{c c c c c c}\hline
Model & 2B & B& H		&2H\\
\hline\hline
2B & --	&0.63	& 1.28& 2.17\\
B  &		& --	& 1.74& 1.89\\
H  &		&			& --	&  1.23\\
\hline
\end{tabular}
\end{center}
\end{table}

%\begin{table}[!htb]
\begin{table}[t]
\caption{$\delta \mathrm{log} (p_1)$, where $p_1$ is measured in
$\mbox{dyn}\,\mbox{cm}^{-2}$, 
in broadband (burst-optimized) noise $\times \left(
D_{\text{eff}}/ 100\,\mbox{Mpc}  \right)$. Compare the range in $p_1
$ of the candidates from Table~\ref{tab:modprop} and Fig.~\ref{fig:candidates}.\label{tab:deltap}}
\begin{center}
\begin{tabular}{c c c c c c}
\hline
Model & 2B & B & H&2H\\
\hline\hline
2B & --&0.10&  0.20& 0.32\\
B & & --&  0.25& 0.26\\
H   & & & --&  0.17\\
\hline
\end{tabular}
\end{center}
\end{table}

\subsection{Some sources of error}

We have cavalierly neglected many higher order but likely relevant 
effects in this preliminary analysis.

It is possible that tidal effects measurably influence the orbital evolution
before the start of the numerical simulations, as estimated in
\cite{FlanaganHinderer2007}, slowly enough not to be seen over the few
cycles of the waveform matched to PP in this analysis. In one sense this
analysis is a worst-case scenario, as it assumes exact PP behavior before
the numerical match.  Earlier drift away from point particle dynamics would
give larger differences between waveforms, and more sensitive radius
measurement, but poses a challenge by requiring accurate numerical
simulation over many cycles to verify EOS effects. Combining numerical
estimation with PN analyses like those of \cite{FlanaganHinderer2007}
and/or quasiequilibrium sequence information may clarify the transition
between effectively PP and tidally influenced regimes.

The waveforms themselves have some residual eccentricity from initial data
and finite numerical resolution. One can estimate eccentricity error by
comparing the plus polarization to the cross polarization shifted by
$\pi/2$ from the same numerical waveform. This results in
$\overline{\varrho_{\text{diff}}}$ of $\sim 0.3$ for HB and 2H, rather than
the expected quadrupole polarization cross-correlation of zero.  (The other
waveforms produce $\overline{\varrho_{\text{diff}}} \sim 0.05$ between
polarizations).

The length of the inspirals, as discussed in the section on PN matching,
limits precision in choosing the best match time.  We can estimate these
effects on current results by varying the match region considered; this
changes $\overline{\varrho_{\text{diff}}}$ results by up to $\sim 0.5$ at
100\,Mpc in the broadband detector. The resolution from existing numerical
simulations is thus comparable to the difference between parameters of the
three closest models.

We have only a coarsely sampled family of waveforms; estimates of $\partial
h/\partial \theta$ are limited by this. The value of
$\overline{\varrho_{\text{diff}}}$ for HB-B and H-HB should be half that of
H-B, but instead they are the same or greater---we are hitting the limit of
numerical and matching accuracy.  We can also estimate the validity of the
linear parameter dependence in the central models by comparing HB to
$\left( \mbox{H} +\mbox{B} \right)/2$. This results in
$\overline{\varrho_{\text{diff}}} \simeq  0.8$, another estimate of
systematic error.  

We conclude that, although these are of course first estimates, they should
be better than order-of-magnitude; the uncertainty in each
$\overline{\varrho_{\text{diff}}}$ is comparable to the
$\overline{\varrho_{\text{diff}}}$ between the H/HB/B waveforms, and
smaller than the $\overline{\varrho_{\text{diff}}}$ for larger differences
in EOS and for comparison with PP.

Finally, we note that use  of a Fisher matrix estimate of parameter
measurement accuracy is fully valid only in high SNR limit of
$\overline{\varrho_{\text{diff}}} > 10$ \cite{vallisneri2007pe}, i.e.,
for distances $\lesssim 20$\,Mpc in the broadband detector. 
The results do not take into account multiple detectors, nor multiple
observations, nor parameter correlations.  A full estimation of
EOS parameter measurability will require more detailed analysis, with a
larger set of inspiral simulations sampling a broader region of parameter
space, e.g., with mass ratios departing from unity, and with several more orbits
prior to merger.

\section{Conclusions}

Gravitational wave astrophysics provides a promising new window on behavior
of cold dense matter.  We estimate that realistic EOS will lead to
gravitational inspiral waveforms which are distinguishable from point
particle inspirals at an effective distance of 100\,Mpc or less in a
burst-optimized Advanced LIGO configuration, as good or better than in 
narrow band detector configuration.

While the standard noise configuration of Advanced LIGO is not sensitive to
the differences in the waveform, the preliminary standard noise curve of
the Einstein Telescope indicate the ability to differentiate between EOS at
roughly double the distance as broadband Advanced LIGO\@. In general,
detuning detectors to be more sensitive at frequencies above 700\,Hz will
lead to improved gravitational wave constraints on neutron star EOS and
radius. 

First estimates of parameter measurability in broadband Advanced LIGO show
$\delta R \sim 1\,\mbox{km} \times (100\,\mbox{Mpc}/D_{\text{eff}})$. One
can also consider a direct constraint on the EOS pressure parameter $p_1$
at a rest mass density $\rho_1 = 5 \times 10^{14}$\,g\,cm$^{-3}$  of
$\delta p_1 \sim 10^{32}\,\mbox{dyn}\,\mbox{cm}^{-2}$ at an effective
distance $D_{\text{eff}}=100\,\mbox{Mpc}$.  These estimates neglect
correlations between these parameters and other details of internal
structure, but such details are expected to give relatively small
corrections to the tidal effects.

While our results must still be considered preliminary, they strongly
motivate further work on gravitational wave constraints from binary neutron
star inspirals.  Future numerical simulations with longer inspirals and
increased coverage of parameter space should improve the accuracy of the
estimates.

\acknowledgments

This work was supported in part by NSF grants PHY-0503366, PHY-0701817 and
PHY-0200852, by NASA grant ATP03-0001-0027, and by JSPS Grants-in-Aid for
Scientific Research(C) 20540275 and 19540263. CM thanks the Greek State
Scholarship Foundation for support. Computation was done in part in the
NAOJ and YITP systems.

\bibliography{paper}
\end{document}